1  **4D imaging of fracturing in organic-rich shales during heating**

3  *Maya Kobchenko[1], Hamed Panahi[1,2], François Renard[1,3], Dag K. Dysthe[1], Anders Malthe-*

4  *Sørenssen[1], Adriano Mazzini[1], Julien Scheibert[1,4], Bjørn Jamtveit[1] and Paul Meakin[1,5,6]*

6  [1] Physics of Geological Processes, University of Oslo, Norway

7  [2] Statoil ASA, Oslo, Norway

8  [3] Institut des Sciences de la Terre, Université Joseph Fourier and CNRS, Grenoble, France

9  [4] Laboratoire de Tribologie et Dynamique des Systèmes, CNRS, Ecole Centrale de Lyon, Ecully CEDEX,

10  France.

11  [5] Idaho National Laboratory, Idaho Falls, USA

12  [6] Institute for Energy Technology, Kjeller, Norway

**Abstract**

To better understand the mechanisms of fracture pattern development and fluid escape in low permeability rocks, we performed time-resolved *in situ* X-ray tomography imaging to investigate the processes that occur during the slow heating (from 60° to 400°C) of organic-rich Green River shale. At about 350°C cracks nucleated in the sample, and as the temperature continued to increase, these cracks propagated parallel to shale bedding and coalesced, thus cutting across the sample. Thermogravimetry and gas chromatography revealed that the fracturing occurring at ~350°C was associated with significant mass loss and release of light hydrocarbons generated by the decomposition of immature organic matter. Kerogen decomposition is thought to cause an internal pressure build up sufficient to form cracks in the shale, thus providing pathways for the outgoing hydrocarbons. We show that a 2D numerical model based on this idea qualitatively reproduces the experimentally observed dynamics of crack nucleation, growth and coalescence, as well as the irregular outlines of the cracks. Our results provide a new description of fracture pattern formation in low permeability shales.

## 1. Introduction

A wide variety of geological phenomena involve the generation and migration of fluids in low permeability rocks. For example, dehydration of sediments in subduction zones generates large fluxes of water that rise along low-permeability subduction interfaces, and provide a mechanism for creep and/or slow earthquakes [*Obara*, 2002]. Similarly, the illitization of clays at depth and the production of methane in organic-rich shales were suggested to contribute to the development of overpressure and the formation of piercement structures, which are manifest on the surface as mud volcanoes [*Mazzini et al,* 2009]. Also, the emplacement of magmatic bodies into sedimentary basins induces rapid decomposition of organic matter, and the resulting gases migrate through low permeability rocks in quantities that may significantly alter the climate [*Svensen et al.,* 2004]. In all these geological systems, the migration of a fluid through low permeability rocks is coupled with deformation. This type of coupling between fluid flow and deformation is also important in the primary migration of hydrocarbons.

Primary migration is the transport of hydrocarbon fluids from extremely low permeability source rocks, in which they are generated, to more permeable rocks through which they migrate to a trap (reservoir) or to the surface. This natural process is of both economic and fundamental interest. As the organic-rich fine grained sediment from which the source rock is formed is buried, the organic material is transformed into complex high molecular weight/cross-linked organic oil and gas precursors (kerogen). As the burial depth increases, the temperature and pressure rise, and kerogen decomposes into low molecular weight hydrocarbon fluids (gas and oil), which have much lower viscosities than the kerogen. The generated hydrocarbon fluids escape from the shale into secondary migration pathways, by processes that remain enigmatic, in spite of decades of investigation [*Bjørlykke, 2010*].

Fracturing is commonly cited as a likely mechanism to increase the permeability of source rocks and provide pathways for the generated hydrocarbons [*Berg and Gangi*, 1999; *Lash and Engelder*, 2005]. During kerogen decomposition, generation of less dense fluids leads to pore-pressure build-up, which may cause fracturing of the host rock. The presence of microcracks is commonly observed in thin sections of recovered source rock samples [*Capuano*, 1993; *Vernik, 1993; Marquez and Mountjoy*, 1996; *Lash and Engelder, 2005*], suggesting that microcracks may be involved in fluid migration.

Under natural conditions, this fracture process takes place at depths of several kilometers over millions of years, making its monitoring impossible. Therefore, it is very important to construct adequate models of primary migration. Several theoretical and numerical studies of fracture formation in organic-rich shales have been described in the scientific literature [*Ozkaya, 1988; Jin et al.,* 2010]. *Jin et al.* [2010] introduced a fracture mechanics model of subcritical crack propagation and coalescence, based on the assumption of linear elastic behavior of the rock. Although the model provides an estimate of the fracture propagation time, the 3D geometry and mechanism of fracture formation in heterogeneous shale remain unclear.

To better understand the complex phenomenon of hydrocarbon expulsion from very low permeability source rock, we employed a new experimental approach – real-time 3D X-ray tomography – during gradual heating of a shale sample coupled with thermogravimetry and petrography analyses before and after heating. This study included only unconfined samples, but it is an important step towards the goal of unraveling the numerous and complex mechanisms controlling primary migration.

**2. Characterization of Green River Shale samples**

The samples were obtained from an outcrop of the organic-rich R-8 unit of the Green River Shale Formation (Piceance Basin, northwestern Colorado, USA). The formation consists of Eocene lacustrine marl/silt sediments [*Ruble et al.,* 2001] with well-developed lamination and anisotropic mechanical properties [*Vernik and Landis*, 1996]. It contains organic matter (total organic content = 9.92 wt% in analyzed samples) present in the form of patches of kerogen, preferentially distributed parallel to the bedding (Figure 1A, B). This part of the formation has never been buried deeply enough to reach temperatures sufficient to cause significant thermal maturation.

Cylindrical core samples (5mm height, 5mm in diameter) cut perpendicular to the bedding were prepared for X-ray tomography. Thin sections were taken before and after heating. Optical microscope images (Figure 1A, B) highlight the micro-fabric of the uncooked shale consisting of alternating lamellae of: coarser and lighter colored carbonate cemented layers rich in siliciclastic grains (e.g. quartz) and pyrite framboids; and darker and finer grained clay-rich layers, containing greater amount of discontinuous organic matter lenses.

**3. 4D imaging and data processing**

Time-lapse X-ray tomography 3D imaging of the shale samples was carried out using beamline ID19 at the European Synchrotron Radiation Facility in Grenoble, France. This non-invasive imaging technique measures the absorption of X-rays, to produce a 3D attenuation map.

The prepared shale cylinder was placed in a home-built furnace in contact with air, with no confining pressure, and it was gradually heated *in-situ* from 60°C to 400°C at approximately 1°C/min. 68 3D scans of this sample were taken, of which 28 were acquired during the heating phase and the rest at 400°C. For all scans, 1500 radiographs were acquired while the sample was rotated over 180°. The beam energy of 20 keV allowed images to be captured with a spatial

100  resolution of 5 µm (5x5x5µm$^3$ voxels) and a time resolution of 11-14 minutes per 3D scan. 3D

101  raw-tomograms representing the microporous structure of the sample at different stages of

102  heating were constructed from the X-ray data. The final volumes contained 830$^3$ voxels coded in

103  8-bit gray levels.

104  The 3D images were processed in two ways. First, the strain field was measured in a

105  typical 2D vertical slice of the volume, using digital image correlation [*Rechenmacher and*

106  *Finno, (2004); Hild and Roux*, 2006*; Viggiani, (2009); Bornert et al, 2010*]. Second, in order to

107  determine crack geometries and track crack propagation, the shape, volume and morphology of

108  the cracks were analyzed in 3D using the AvizoFire© software package. Quantitative analysis of

109  the crack formation required isolation of the cracks from the rock matrix. Due to the small crack

110  opening (4-5 voxels), the following procedure was applied: firstly a binary mask was used to

111  delete the background; then an edge-preserving smoothing filter based on Gaussian smoothing

112  combined with a non-linear diffusion algorithm was applied; then a "watershed" procedure

113  enabled individual cracks to be isolated [*Sonka et al*, 1999]. The final result of this segmentation

114  procedure was a series of cracks consisting of connected voxels, marked by different

115  labels/colors (Figure 2A).

116  **4. Observation of deformation and crack formation**

117  Correlation analysis and 3D image analysis were performed to determine the deformation

118  of the sample before fracturing and the geometries of the developing cracks. The spatial

119  fluctuations of the attenuation maps (see Figure 2B) served as markers when the digital

120  correlation technique was used to compare successive images. To calculate the correlation matrix

121  we used a correlation box of size 25 by 25 pixels (125 by 125 µm$^2$), which enabled us to measure

122  the spatial distribution of micro-displacements.

Between room temperature and 300°C, the shale dilated anisotropically in the vertical (perpendicular to the shale bedding) and horizontal (parallel to the shale bedding) directions, and the strain curves showed a quasi-linear increase with temperature, as expected for linear thermal expansion (Figure 3A). The coefficient for thermal expansion was determined to be $5.5*10^{-5}$/°C in the vertical direction and $2.5*10^{-5}$/°C in the horizontal direction, which clearly indicates the anisotropy of the shale, in agreement with other studies on the same shale [*Grebowicz*, 2008]. At 300°C, the vertical expansion started to deviate from linearity, which is likely related to the onset of organic degassing before crack formation. At a temperature of about 350°C the sample undergoes rapid localized deformation owing to fluid generation and the onset of fracturing. The sample breaks, black structures corresponding to the newly formed cracks appear in the images (see Figure 2B), which have no equivalent in the preceding ones, thus ruining the correlation technique. Moreover, after the sample fractures, global movement of the sample occurs (translation and rotation), with displacements from one data set to the next of amplitude greater than 12 pixels, which was estimated to be the maximum displacement that can be accurately measured using the correlation technique. Even though the correlation results during and after fracturing are not trustworthy, this technique accurately determines the sample deformation before fracturing occurs.

With 3D tomography, we imaged 3 stages of fractures propagation. The first microfracture pattern was detected at a temperature of about 350°C. Figure 2A shows a 3D rendering of the most opened fractures at T=391°C (third time step) and Figures 2B and 2C show a vertical slice through the tomography image. The general direction of crack propagation followed the shale bedding, and no perpendicular fractures were observed (Figure 2B). Pyrite

145   grains, observable as bright spots in the tomography images (Figure 2C), affected crack growth
146   by pinning the crack front, and controlling the out-of-plane fluctuations of the crack path.

147   The fractures can be described in terms of two rough, essentially parallel surfaces that enclose the fracture aperture. Both surfaces can be described by height functions, $h_1(x,y)$ and $h_2(x,y)$, where $(x,y)$ is the position in the common plane. Figure 4A shows the topography of fracture surface $h_1(x,y)$, where the fluctuation of surface function $h_1(x,y)$ around the flat plane $(x,y)$ is indicated by the color code. Cracks have essentially constant aperture widths - the thickness function $h_2(x,y) - h_1(x,y)$ (4-5 pixels, i.e. 20-25 micrometers, see Figure 4B), and rough irregular outlines. The amplitude of the topography variation is around 10-15 pixels (i.e. 50-75 micrometers), as seen in the Figure 4A. The mid-plane ($H(x,y) = [h_1(x,y) + h_2(x,y)]/2$) is also rough, and it fluctuates about a flat plane – the "plane of the fracture". The projections of $h_1(x,y)$, $h_2(x,y)$ and $([h_1(x,y) + h_2(x,y)]/2$ into this plane, in a direction perpendicular to the plane, have a common shape, which consists of a continuous region with rough edges. When one fracture coalesces with another, a new fracture is formed, and it can be described in terms of the rough surfaces $H^{new}(x,y)$, $h_1^{new}(x,y)$ and $h_2^{new}(x,y)$. The projections of these rough surfaces are also a common shape, which consists of a continuous region with rough edges. The planes of the coalesced fracture and the two fractures that coalesced to form it are more-or-less parallel, and also quasi-parallel to the shale lamination. Viewed from "above", in the direction perpendicular to the plane of the fractures and the lamination, the fractures are continuous, and they grow and coalesce in three successive stages (Figure 5A). As the temperature rose, cracks nucleated (1), grew and coalesced in a quasi-static manner (2) until they spanned the sample (3). The growth of the surface area of the biggest fracture with increasing temperature is shown in Figure 3D.

**5. Organic decomposition induces fracturing in the shale**

168    Thin sections were studied in order to compare petrographic characteristics of the shale
169 before and after heating. Before heating, organic precursors, which were preferentially oriented
170 parallel to the shale bedding, could be distinguished throughout the sample (see Figure 1A, B).
171 After heating, an abundance of cracks, partially filled with residual organic material, was
172 distributed parallel to the bedding (Figure 1C, D). Petrographic observations revealed that cracks
173 propagated mainly in the finer grained layers where the highest concentrations of organic matter
174 lenses were observed (Figure 1A, C). The coarser grained layers, where quartz grains and pyrite
175 framboids were present in higher concentration, also displayed better cementation with larger
176 calcite crystals. The preferential location of fracture propagation is ascribed to two main factors:
177   a) Higher amounts of organic matter (kerogen lenses), which decomposes leading to fluid
178      formation, internal pressure build up and eventually fracture initiation and propagation.
179   b) Finer grained intervals are less cemented than the coarser grained ones and they fracture
180      more easily.

181 The link between hydrocarbon generation and fracturing was tested using thermogravimetry and
182 gas chromatography. Aerobic and anaerobic (nitrogen) thermo-gravimetry analyses on 500 mg
183 samples were performed to investigate the presence of organic and inorganic carbon (carbonates)
184 using a ATG/SDTA 851 Mettler Toledo apparatus. We monitored mass loss of the sample during
185 heating at 10°C/min in air or nitrogen between 20°C and 1000°C. The loss of mass during
186 heating occurred in distinct stages (Figure 3B), and the temperature range of each stage indicates
187 the nature of the component that evaporates. We also used gas chromatography (GC 5890-MS
188 5973 Agilent) to analyze the gas that escaped (water, $CO_2$, and organic volatiles) during heating
189 at a rate of 5°C/min in air. The first step of mass loss (Figure 3B) in the temperature range 300 -
190 450°C corresponds to the release of various organic molecules (alkanes, alkenes, toluene,

191 xylenes), water and $CO_2$ (first peak on $CO_2$ emission plot) and indicates decomposition of
192 organic matter. The second step of mass loss around 600 - 800°C indicates decomposition of
193 carbonates. A similar behavior was observed when nitrogen was used instead of air with two
194 peaks of $CO_2$ release located at the same temperatures.

195 Figure 6 shows the correlations between strain evolution, mass loss, $CO_2$ emission and
196 fracture surface area growth. Comparing the temperatures of degassing, mass loss and
197 hydrocarbon release with the temperature of fracturing onset, we conclude that fracturing was
198 induced by overpressure in the sample caused by organic matter decomposition.

199 **6. Discrete model of crack propagation**

200 Based on the experimentally observed fracturing behavior (Figure 5A), a two
201 dimensional (2D) model of in-plane crack nucleation and growth due to internal pressure
202 increase was developed (Figure 7). The choice of a 2D model is based on the observation of
203 planar mode 1 cracks that follow the layering of the shale and the focus on in-plane dynamics
204 rather than the placement of cracks and relation between cracks normal to their plane. To
205 reproduce key characteristics of the crack growth process including the merging of small cracks
206 and the very irregular crack shapes, we used a statistical fracture model [*Alava et al., 2006*]. The
207 main parameter needed is the variability of local strength, and the model neglects long range
208 effects of the stress field. The model includes neither the kinetics of kerogen decomposition nor
209 the elastic properties of the source rock, and therefore it cannot quantitatively predict threshold
210 temperatures and pressures, or rates of primary migration. The objective of the model is to find
211 the minimum number of features that explain the observed behavior. The macroscopic fracture
212 threshold and kerogen transformation kinetics are important to predict at what temperature and
213 rate these phenomena occur, but they do not help us understand how the fracture evolves. The percolation like evolution of the fracture, as opposed to the nucleation of a single fracture that propagates rapidly through the shale, which would be predicted by a macroscopic elastic model with a distributed pressure build-up, shows that the material disorder may cause the slow formation of percolating fluid pathways for expulsion of hydrocarbon fluids at pressures much smaller than those needed to completely fracture the shale. The model focuses on a layer of shale that fractures more rapidly than nearby layers because it has a higher kerogen content. The layer is modeled by a regular square lattice in which every site represents a small organic-rich shale element. Each site is characterized by a randomly assigned breaking threshold $\sigma_{c,i}$. The pressure in the lattice rises incrementally during each time step and, when it exceeds the breaking threshold ($p_i > \sigma_{c,i}$), the site fractures to represent either nucleation of a new crack or growth of a pre-existing crack. This relaxes the stress, which is distributed equally to the nearest non-broken neighbors (long-range elastic interactions are neglected), bringing them closer to failure. The distribution of stress was implemented by reducing the breaking threshold, $\sigma_{c,i} \rightarrow \sigma_{c,i} - d\sigma$, for all nearest non-broken neighbors. Dimensionless units were used for the pressure and breaking thresholds. Each crack is represented by a cluster of broken sites. As soon as sites belonging to different clusters become adjacent, both clusters are merged to form a single crack, and all the merged sites are given the same label/color.

Figure 5B shows three successive snapshots during a simulation. In the early stage, the system contains many small independent cracks. Each crack has a rough front, and over time, individual cracks grow slowly and merge until the whole plane is covered. Figure 3C displays the increase of the area of the largest final simulated crack. Crack growth occurs in three stages: (1) the microcracks are all separated and their surface areas grow gradually; (2) the cracks start to coalesce, the rate at which the fractured area increases accelerates, and growth in the total fracture area is dominated by distinct jumps and (3) ultimately the system is dominated by one large fracture, with an area that grows by intermittent increases (see also the corresponding snapshots 1-3 in Figure 5B). The qualitative trend of the fracture area growth in the simulation (Figure 3C) is similar to that observed in the experiment (Figure 3D).

The model is similar to the fiber bundle model with local load sharing, which has been intensively studied model for material failure [*Pradhan,* 2010]. We suggest that it can be applied to other geological systems in which chemical reactions induce volume increase and stress build-up in rocks. These systems are widespread and include, apart from primary migration of hydrocarbons, weathering of rocks near the surface [*Røyne et al*., 2008] and dehydration of serpentines in subduction zones [*Jung et al.,* 2004].

**7. Discussion and Conclusion**

Time-resolved high-resolution synchrotron X-rays tomography was performed during gradual heating (from 60 to 400ºC) of organic-rich immature shales. At 350 ºC the nucleation of many small cracks was detected. With further temperature increase these microcracks propagated parallel to the shale bedding, coalesced and ultimately spanned the whole sample.

The central point of our work is the observed correlation between hydrocarbon expulsion and fracturing within the sample. To do this, we combined thermogravimetry, gas chromatography, strain analysis and 3D observation of fracture formation and analyze the data in 3 steps:

- Analyzing tomograms, we found that that fracturing begins at a temperature of about 350°C.

- Using thermogravimetry, we determined that the sample started to loose mass in the same temperature range (about 350°C). This alone does not provide us with the composition of the lost material.
- Using gas chromatography, we determine that the gases expelled at temperatures near 350°C were mainly oxidized hydrocarbons, and we conclude that they originate from kerogen decomposition.

In the literature, some shale rocks are reported to contain horizontal in the direction of shale bedding (mode I) fractures as well as vertical (perpendicular to the shale bedding) fractures. The presence of vertical fractures indicates that the maximum stress is vertical [*Olson, 1980; Smith, 1984*]. However at shallow depths, where the magnitudes of both the vertical and horizontal stresses are similar, horizontal hydro-fractures of significant lateral extent can be created due to anisotropy of shales [*Thomas, 1972; Jensen, 1979*]. Horizontal fractures are also observed to develop in clay-rich shales in response to high overpressures during maturation [*Littke, 1988*], even in regions where the vertical stress is larger than the horizontal stress. The reason for this is thought to be the strong lamination of these shales [*Lash and Engelder, 2005; Vernik, 1993*]. Theoretical studies [*Ozkaya, 1988*] also showed that vertical cracks are unlikely to form by oil generation and the excess of oil pressure is sufficient to cause lateral fracturing if the aspect ratio of kerogen patches is sufficiently large.

In nature petroleum generation takes place in the 80-150°C temperature range, and this occurs typically over time periods of 1-100 million years (up to 500 million in some geological situations). The heating rate for a basin like the North Sea is 1-2°C/million years, but during rapid sedimentation and subsidence the heating rate is as high as 10°C/million years [*Bjorlykke,*

281  *2010*]. In our experiments, conducted in a much shorter time, fracturing was observed to occur in
282  the temperature range of 300-400°C.

283  The decomposition of kerogen involves a very large number of coupled chemical
284  reactions, and a temperature dependent rate constant is associated with each reaction. In general,
285  the rates of these reactions can be represented approximately by the Arrhenius function $K_i = A_i T^n exp(-E_i^a/RT)$,
286  where $K_i$ is the rate constant for the $i^{th}$ reaction, $A_i$ is its frequency factor, $E_i^a$ is
287  its activation energy, $R$ is the gas constant and $T$ is the absolute temperature. In most cases, $n$ is
288  small, the temperature dependence is dominated by the exponential term and the Arrhenius
289  equation is often expressed in the form $A_i' exp(-E_i^a/RT)$. The overall rate of the organic maturation
290  process, $K$, can also be represented by a similar Arrhenius function $K=Aexp(-E^a/RT)$, although
291  the nature of the decomposition products is also somewhat temperature dependent. If the
292  activation energy, $E^a$, is large enough, the maturation rate increases rapidly with increasing
293  temperature, and this explains why a higher temperature can be used to accelerate the maturation
294  process and achieve a maturation time that is short enough to conduct laboratory experiments.

295  Under natural conditions cracks form during maturation at lower temperature, and
296  because of the low rate of pressure buildup crack growth is probably a slow subcritical fracturing
297  process. Based on theoretical calculations by [*Ozkaya, 1988*], we can estimate the conditions for
298  the initiation of lateral cracking. [*Ozkaya, 1988*] considered a linearly elastic and isotropic source
299  rock containing isolated kerogen flakes enclosed in an impermeable shale matrix, and showed
300  that the aspect ratio of the kerogen particles may influence the initiation of horizontal
301  microcracks in the source rock beds under maximum vertical principal stress. Initiation of
302  microcracks as a function of kerogen flakes shape occurs when:

303  $$\Delta P\left(\frac{2w}{h}-1\right) > S_v(2-k)+C, \qquad [1]$$

where $S_v$ is the principal vertical stress, $k$ is the lateral to vertical stress ratio, $C$ is the strength of the rock, $\Delta P$ is the excess oil pressure induced during kerogen decomposition, $w$ is the length of kerogen particle, and $h$ is the thickness of kerogen particles. For example, for the kerogen patch shown on the Figure 1C, the aspect ratio is $w/h=50\mu m/3\mu m\approx16$. For $k=0.75$, $\Delta P=5$Mpa, $C=10$MPa [*Ozkaya, 1988*], equation [1] indicates that lateral fracturing may occur if the vertical stress $S_v$ is lower than 116 MPa, which corresponds to a depth $h$ smaller than 4000 m.

One limitation of this calculation is the assumption that the kerogen patches are surrounded by impermeable rock. In reality, shales have very small permeabilities and the transformation reaction is slow. The generated fluids may migrate through the rock matrix without producing pressure gradients that are large enough to cause fracturing. However magma emplacement in sedimentary basins may result in much higher maturation rates and the rapid production of hydrocarbon fluids may cause rapid internal pressure build up and fracture generation.

To summarize our observations, correlation analysis monitoring of the deformation of the sample indicated abrupt swelling perpendicular to the bedding just before the cracks formation. Petrography observations showed that the main cracks initiate in the finer grained clay-rich layers where a higher amount of organic matter is present. Thermogravimetry and gas chromatography analyses showed that the sample began to lose significant mass and release water, $CO_2$ and hydrocarbon gasses at about 350°C. These observations support a scenario in which the kerogen, present in thin lenses, starts to decompose around 350ºC, causing volume increase and internal pressure build up leading to fracturing. The fracturing mechanism observed experimentally, including crack formation and crack geometry, was successfully reproduced by a 2D fiber-bundle lattice simulation. The success of the statistical fracture model indicates that material disorder and local elastic interactions play a key role in the development of the observed percolation like fracture. The implication is that percolating low permeability fluid pathways are formed in the shale long before macroscopic fracturing occurs.

In the present study we do not characterize fracturing of source rocks under natural conditions. In particular, no confinement was applied to the sample. We therefore suggest that the results of our experiments can be directly relevant to better understand induced fracturing of shales located at shallow depth or outcrops, where the confinement pressure is low. More generally, the methods developed here will be relevant for future studies under more realistic conditions.

**Acknowledgments:** We acknowledge support by the Petromaks program of the Norwegian Research Council. We thank Elodie Boller at ESRF for her help during the tomography scans and Rodica Chiriac for her help with the thermogravimetry analyses.

341 **References**

342 Alava, M. J., Nukalaz P. K. V. V., and S. Zapperi (2006), Statistical models of fracture,
343 *Advances in Physics, 55, 349-476.*

344 Berg, R.R., and A.F. Gangi (1999), Primary migration by oil-generation microfracturing in low-
345 permeability source rocks: application to the Austin Chalk, Texas, *AAPG Bull., 83, 727-756.*

346 Bjørlykke, K. (2010), Petroleum Geoscience: From Sedimentary Environments to Rock Physics,
347 *Springer, Berlin, Germany.*

348 Bornert, M., Vales, F., Gharbi, H., and D.N. Minh (2010), Multiscale full-field strain
349 measurements for micromechanical investigations of the hydromechanical behaviour of clayey
350 rocks, *J. Geophys. Res., 46, 33-46.*

351 Capuano, R.M. (1993), Evidence of fluid flow in microcracks in geopressured shales, *AAPG*
352 *Bull., 77, 1303-1314.*

353 Grebowicz, J. (2008), Thermal properties of Green River oil shales, Joint Meeting of The
354 Geological Society of America, Soil Science Society of America, American Society of
355 Agronomy, Crop Science Society of America, Gulf Coast Association of Geological Societies
356 with the Gulf Coast Section of SEPM, 2-10 October 2008, Houston TX, paper 346-1.

357 Hild, F., and S. Roux (2006), Digital Image Correlation: from Displacement Measurement to
358 Identification of Elastic Properties – a Review, *Strain, 42, 69-80.*

359 Jensen, H.B. Oil shale in situ research and development // Talley Energy systems final report, US
360 Department of Energy DOE/ LC/ 01791-T1, 1979.

361 Jin, Z.-H., S. E. Johnson, and Z. Q. Fan (2010), Subcritical propagation and coalescence of oil-
362 filled cracks: Getting the oil out of low-permeability source rocks, *Geophys. Res. Lett., 37,*
363 *L01305.*

Jung, H. (2004), Intermediate-depth earthquake faulting by dehydration embrittlement with negative volume change, *Nature, 428, 6982, 545-549.*

Lash, G.G., and T. Engelder (1992), An analysis of horizontal microcracking during catagenesis: Example from Catskill delta complex, *AAPG Bull., 89, 1433-1449.*

Lash, G.G., and T. Engelder (2005), An analysis of horizontal microcracking during catagenesis: Example from Catskill delta complex, *AAPG Bull., 89, 1433-1449.*

Littke, R., Baker, D.R., and D. Leythaeuser (1988), Microscopic and sedimentologic evidence for the generation and migration of hydrocarbons in Toarcian source rocks of different maturities, *Org. Geochemistry, 13, 549-559.*

Marquez, X.M., and E.W. Mountjoy (1996), Microcracks due to overpressure caused by thermal cracking in well-sealed Upper Devonian reservoirs, deep Alberta basin, *AAPG Bull., 80, 570-588.*

Mazzini, A. (2009), Mud volcanism: Processes and implications, *Marine and Petroleum Geology, 26(9): 1677-1680.*

Obara, K. (2002), Nonvolcanic deep tremor associated with subduction in southwest Japan, *Science, 296, 1679-1681.*

Olson, W.A. (1980), Stress relaxation in oil shale. In: Proc. 21$^{st}$ Symposium on Rock Mechanics, 517-520.

Ozkaya, I. (1988), A simple analysis of oil-induced fracturing in sedimentary rocks, *Marine and Petroleum Geology, 5, 293-297.*

Pradhan, S., Hansen, A. and B. K. Chakrabarti (2010), Failure processes in elastic fiber bundles, *Rev. Mod. Phys., 5, 499-555.*

Rechenmacher, A.L., and R.J. Finno (2004), Digital image correlation to evaluate shear banding in dilative sands, *Geotechn. Testing J., 27, 13-22*.

Røyne, A., Jamtveit, B., Mathiesen, J., and A. Malthe-Sorenssen (2008), Controls of rock weathering rates by reaction-induced hierarchical fracturing, *Earth Planet. Sci. Lett., 275, 364-369*.

Ruble, T. E., M.D. Lewan, and R.P. Philp (2001), New insights on the Green River petroleum system in the Uinta basin from hydrous pyrolysis experiments, *AAPG Bull., 85, 1333 -1371.*

Sandvik, E.I. and J.N. Mercer (1990), Primary migration by bulk hydrocarbon flow. *Adv. In Org. Geochemistry, 16, 83-89.*

Smith, J.W., and Chong, K.P. (1984) Introduction to mechanics of oil shale. In: Chong, K.P., Smith, J.W. (Eds): Mechanics of Oil Shale. Elsevier, 198, 1-41.

Sonka, M., Hlavac, V., and R. Boyle (1999), Image processing, analysis and machine vision, *PWS Publishing, Pacific Grove, CA, USA*.

Svensen, H., Planke, S., Malthe-Sorenssen, A., Jamtveit, B., Myklebust, R., Eidem, T.R., and S.S. Rey (2004), Release of methane from a volcanic basin as a mechanism for initial Eocene global warming, *Nature, 429, 6991, 542-545*.

Thomas, H.E. (1972), Hydraulic fracturing of Wyoming Green River oil shale: field experiments, phase I, *US Bureau of Mines Report Investigation 7596*.

Vernik, L., and C. Landis (1996), Elastic anisotropy of source rocks: Implications for hydrocarbon generation and primary migration, *AAPG Bull., 80, 531-544*.

Viggiani, G. (2009), Mechanisms of localized deformation in geomaterials: an experimental insight using full-field measurement techniques, *Mechanics of Natural Solids, 105-125*.

409 **Appendix A**

410 We provide a code in MATLAB that reproduces the results of Figures 3C and 4B. Note that the

411 MATLAB Image Processing Library is required to run this code.

412

```matlab
% Fiber bundle 2D model with local stress redistribution
L = 100;                        % The layer of shale has a size of LxL sites
sigmac = rand(L,L);      % Random strength thresholds assigned for every site
fractured = zeros(L,L);    % =1 if the site is fractured
dsigma = 0.15;                  % Amount of stress redistribution among the
unbroken neighbors when the site fractures
dp = 0.001;                     % Increment in pressure
pmax = 0.4;                     % Maximum value of pressure
pvalue = (0:dp:pmax);      % Pressure range
nfrac = zeros(length(pvalue),1);     % Number of fractured sites
nlargest = zeros(length(pvalue),1); % Size of the largest fracture
figure(1)
for ii = 1:length(pvalue)
% at each step of the program, pressure rises by the amount dp
   p = pvalue(ii);
   ndo = 1;
   while ndo>0
       ndo = 0;
       i = find(sigmac<p);
% find the site with a breaking threshold lower than the pressure value
       [ix,iy] = ind2sub(size(fractured),i); % find coordinates of this site
       for j = 1:length(i)
           if (fractured(i(j)) == 0)
               fractured(i(j)) = 1;
% change the state of this site to fractured
               nx = ix(j);
               ny = iy(j);
               xneighbors = [nx-1, nx+1, nx, nx];  % select the four
neighbors and find the edges of the system
               yneighbors = [ny, ny, ny-1, ny+1];
               k = find((xneighbors < 1) | (xneighbors > L) | (yneighbors <
1) | (yneighbors > L));
               xneighbors(k)=[];
               yneighbors(k)=[];
               fracturedneighb = ones(1,length(xneighbors));
               for m=1:length(xneighbors)  % find neighbors which are broken
                   if fractured(xneighbors,yneighbors)==1
                       fracturedneighb(m)=0;
                   end
               end
               k=find(fracturedneighb == 0);
% remove neighbors which are broken
               xneighbors(k)=[];
               yneighbors(k)=[];
% distribution of stress among the non-broken neighbors
               if (length(xneighbors) >= 1)
                   dsigmaN = dsigma*4/length(xneighbors);
```

```
                        for m=1:length(xneighbors)
% decrease the strength of non-broken neighbors
                        sigmac(xneighbors(m),yneighbors(m)) =
sigmac(xneighbors(m),yneighbors(m)) - dsigmaN;
                        end
                    end
                    ndo = ndo + 1;
                end
            end
        end
    nfrac(ii) = length(find(fractured>0));    % amount of fractured sites
    [lw,num] = bwlabel(fractured);           % assign a color to every fracture
    img = label2rgb(lw,'jet','k','shuffle');  % label the fractures
    s = regionprops(lw,'Area');              % measure the fracture area
    area = cat(1,s.Area);
    nlargest(ii) = max([max(area),0]);% find the area of the largest fracture
    subplot(1,2,1)
    imagesc(img);                            % plot the image of fractures
    axis equal
    axis tight
    title('Fractures')
    drawnow;
    subplot(1,2,2)
    plot(pvalue(1:ii)/pmax*100,nlargest(1:ii)/L^2*100,'r','linewidth',2);
%area of the largest fracture vs pressure
    xlabel('Pressure,%')
    ylabel('Fracture surface area,%');
    axis([0 100 0 100])
    axis square
    drawnow
end
```

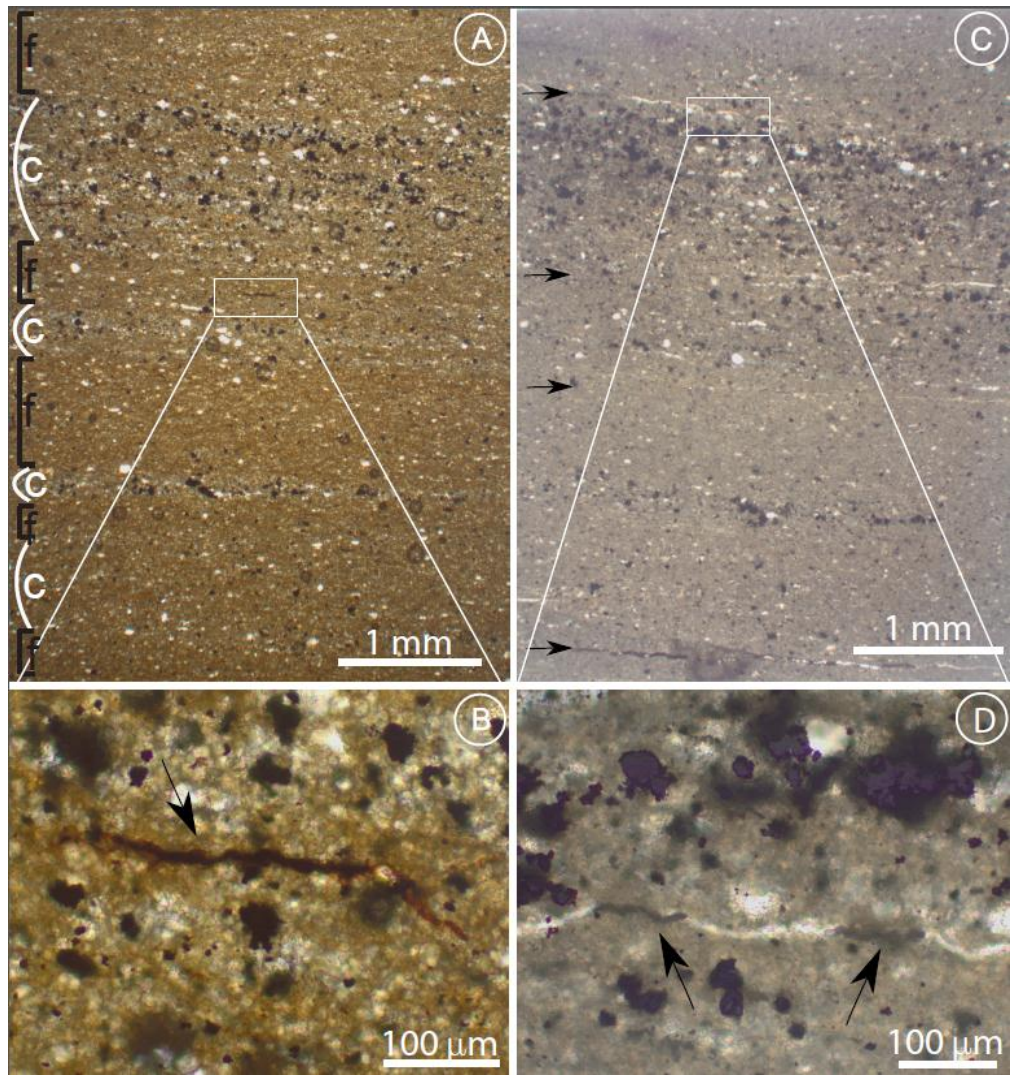

**Figure 1:** Thin section images of Green River Shale sample before and after heating. (A) Interlaminated silt and clay-rich layers before heating. (f) clay-rich layers with higher amount of kerogen lenses, (c) coarser layers with siliciclastic grains. (B) Detail of a kerogen lens. (C) Image of the same sample after heating. Arrows indicate the position of cracks developed during heating. Fractures propagated mainly in the finer grained intervals where the highest concentration of organic matter lenses was also observed. (D) Detail of a crack filled with organic remains (arrows).

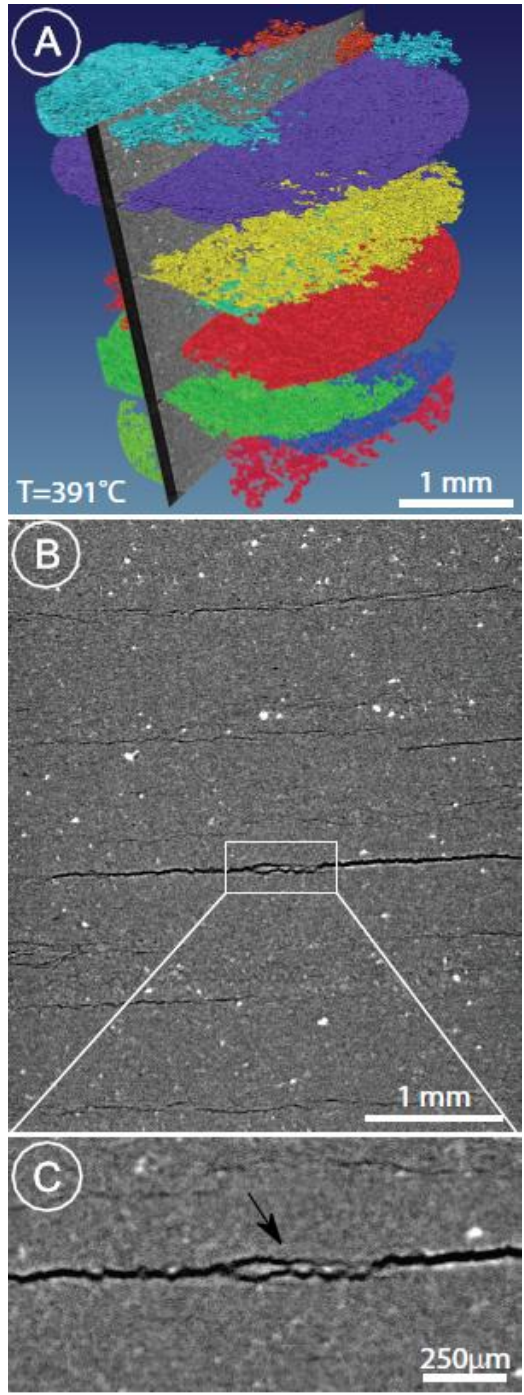

501

502 **Figure 2:** Tomography images of the shale sample at 391°C, corresponding to the moment of
503 maximum crack opening. (A) 3D rendering of final crack network. Each color defines an
504 independent crack. (B) 2D slice showing (dark color) elongated cracks developed parallel to the
505 bedding. (C) Detail of image (B) showing a crack developed around a pyrite grain (arrow).

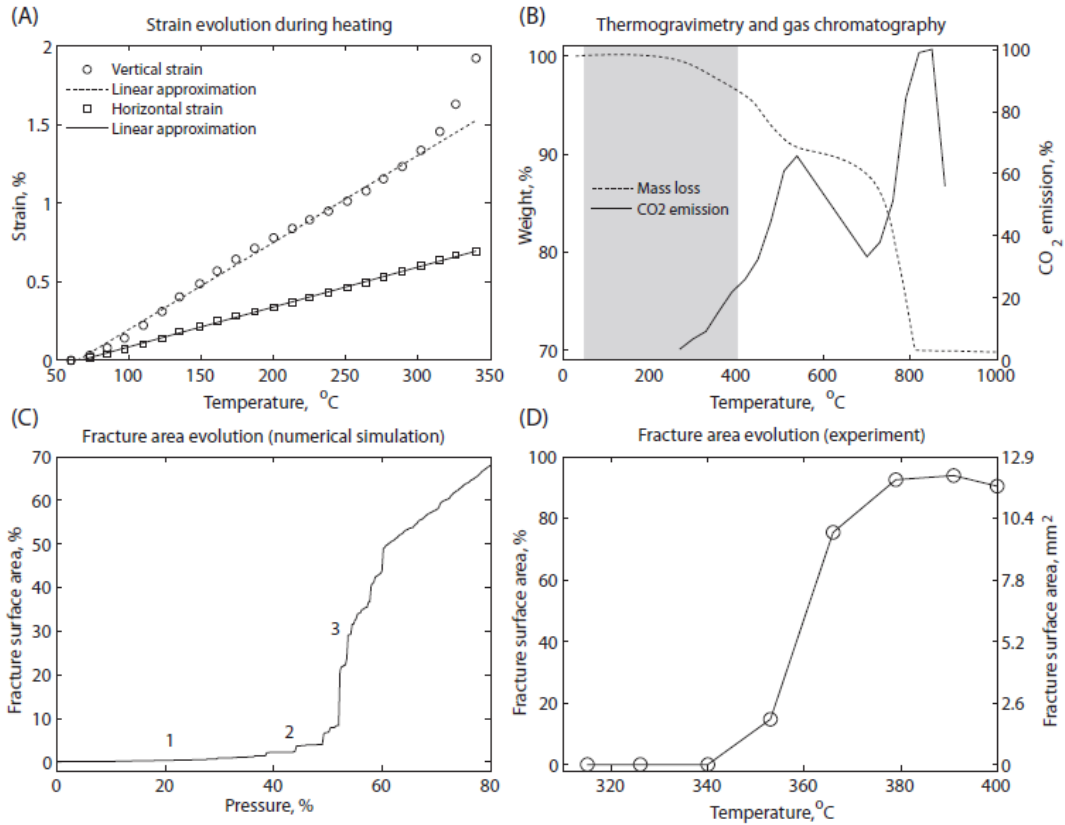

506

507 **Figure 3:** Vertical and horizontal strain in a shale sample during heating. The linear relationship
508 between 70°C and 300°C corresponds to dilation of the sample due to thermal expansion.
509 Because of the anisotropy of the shale, the vertical and horizontal dilations are different. The
510 vertical strain-temperature relationship deviates from linearity at 300°C, at the onset of
511 degassing. Failure of the correlation technique at 350°C corresponds to the fracturing of the
512 sample. (B) Thermogravimetry (mass loss) and carbon dioxide emission analyses in aerobic
513 conditions. The change of slope on the mass loss curve at around 300°C occurs at the same
514 temperature as the onset of $CO_2$ production and corresponds to the onset of kerogen
515 decomposition. The grey shaded area indicates the temperature range of the tomography
516 experiment. The peak of $CO_2$ production and mass loss at around 800°C corresponds to the
517 decomposition of carbonate. (C) Growth of the area of the biggest crack as a function of pressure

(% of maximum applied pressure) in the 2D lattice model. (1-3) – three stages of crack evolution corresponding to nucleation, growth and coalescence (see corresponding snapshots 1-3B in Figure 5). (D) Fracture evolution in the experiment. Growth of the surface area of the largest crack (in % of the sample cross-sectional area and in mm²) as a function of temperature. The slight decrease of fracture surface area observed after 390°C is attributed to partial crack closing after fluid expulsion.

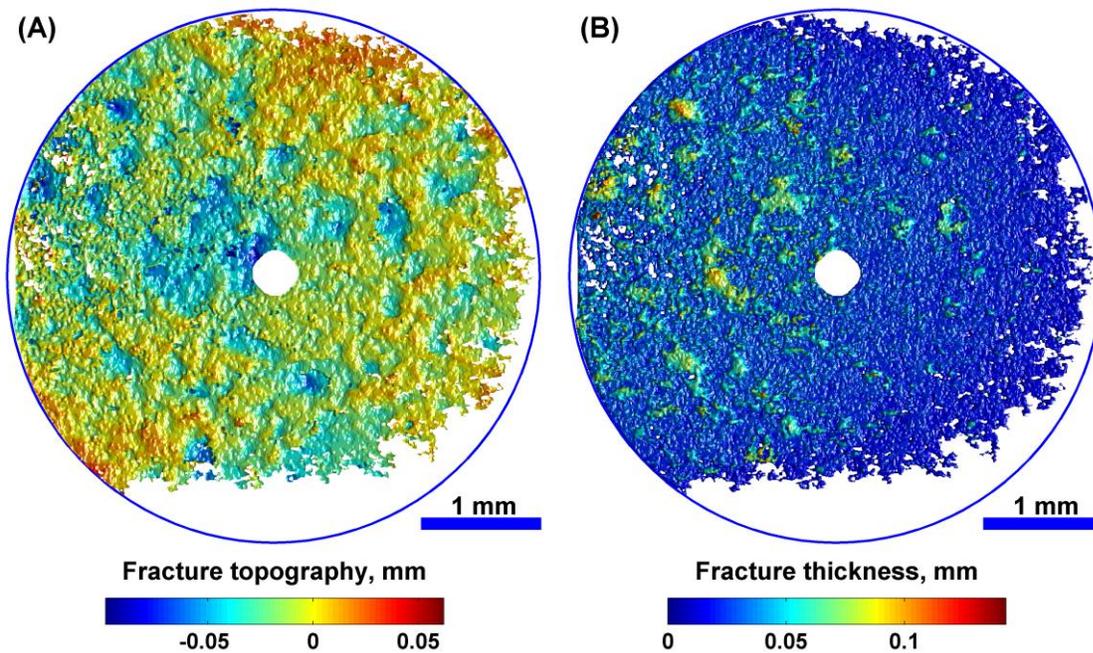

**Figure 4**: Reconstruction of the topography (top) and thickness (bottom) of a fracture extracted from Figure 2A. The circular central region in each image was removed because of a data acquisition artifact. The outer circle defines the boundaries of the sample. (A) Fluctuation of fracture surface height $h_1(x,y)$ around the fracture plane $(x,y)$ is indicated by the color scale. The fracture front is irregular. The topography is created by small heterogeneities (i.e. pyrite minerals) that pin the fracture during its propagation. (B) The fracture thickness, taken as the difference between the upper surface h₁(x,y) and lower surface h₂(x,y) of the fracture, is indicated by the color scale. The thickness is quasi-constant and it is perturbed by pyrite inclusions.

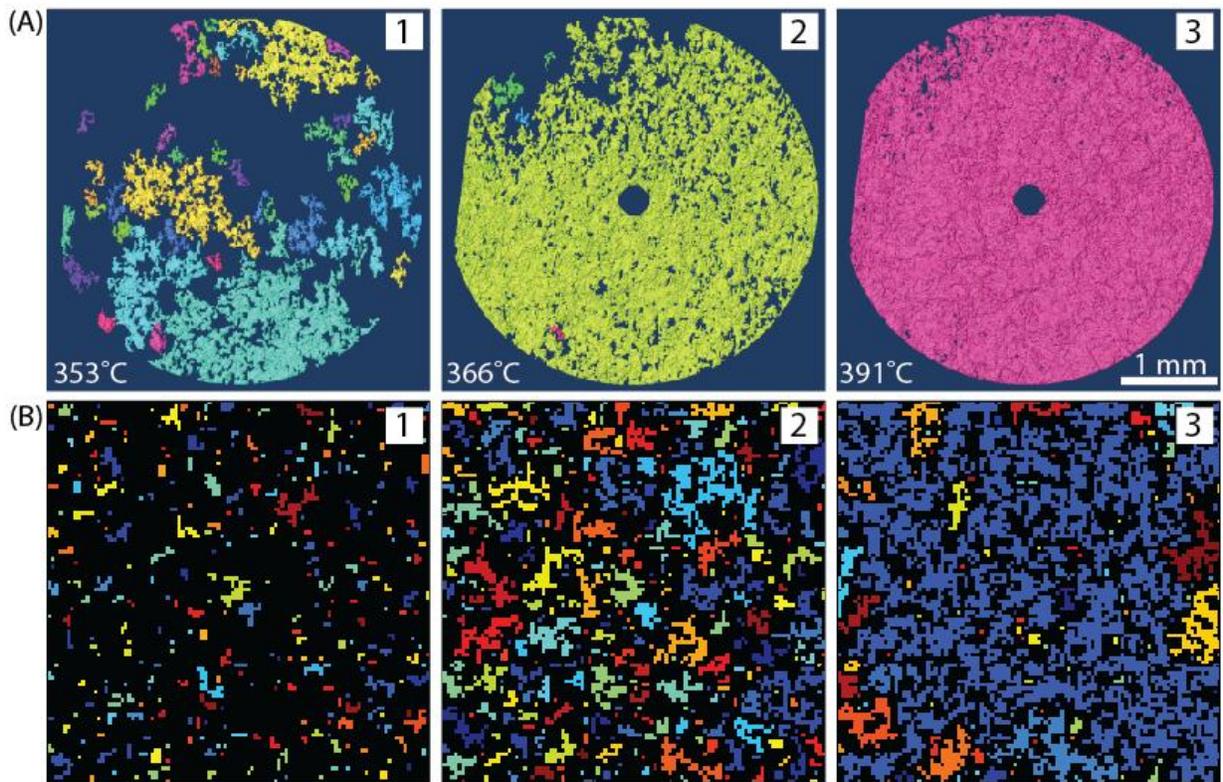

**Figure 5**: Comparison of crack evolution in the experiment and numerical model. (A) Crack propagation dynamics during heating in the experiment. View of the cracks in a kerogen rich layer viewed from a direction perpendicular to the average plane of the cracks. (1) Numerous small cracks nucleated at ~350°C. Each crack is indicated by a different color. (2) Cracks grew and merged with increasing temperature. (3) Ultimately all cracks merged into a single sample-wide crack. The circular central region in each image was removed because of a data acquisition artifact. (B) 2D lattice model at three stages of crack development: nucleation (1), growth (2) and coalescence (3) of cracks (see three stages of the area growth (1-3) in Figure 3C).

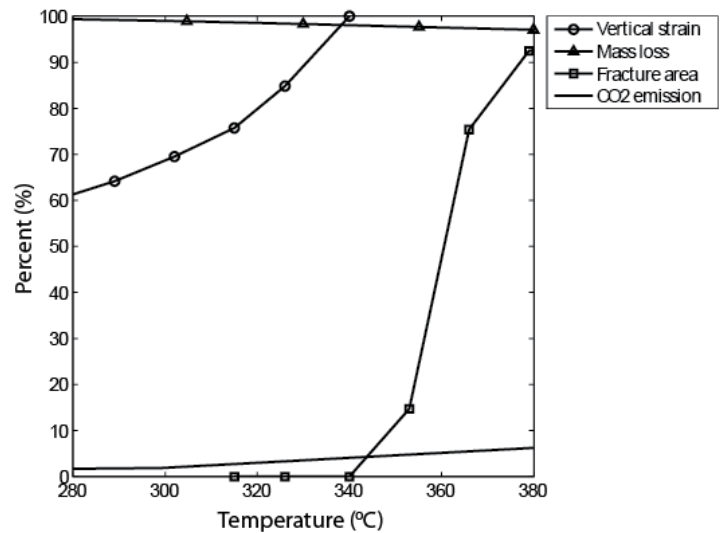

546

547 **Figure 6:** The correlation between vertical strain evolution (perpendicular to the shale
548 lamination) during heating, mass loss, $CO_2$ emission and growth of fracture area in the shale
549 sample. The onset of the mass loss and $CO_2$ emission corresponds to decomposition of organic
550 material. The nonlinear strain growth in the vertical direction, which is caused by internal fluid
551 pressure buildup, leads to the fracturing at 340°C.

552

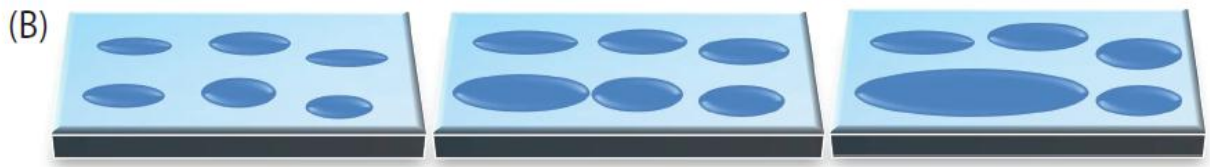

**Figure 7**: Sketch of the 2D discrete model. (A) The rock layer is modeled by a lattice of sites with assigned breaking thresholds. When a site fractures the stress is distributed equally to the non-broken neighbors, making them weaker. Two fractured neighbors are called a continuous crack. (B) The cracks grow by including new neighboring broken sites. When two cracks coalesce they form a bigger crack.